\documentclass{IOS-Book-Article}
\usepackage{mathptmx}
\usepackage[]{graphicx}
\usepackage{amsmath}
\usepackage{algorithm}
\usepackage[noend]{algpseudocode}
\usepackage{balance}
\usepackage{subfig}
\usepackage{tabularx}
\usepackage{multirow}
\usepackage{pdfcomment}
\usepackage{xcolor}
\usepackage{enumitem}

\def\hb{\hbox to 10.7 cm{}}

\begin{document}

\pagestyle{headings}
\def\thepage{}

\begin{frontmatter}              

\title{Power Modelling for Heterogeneous Cloud-Edge Data Centers}

\markboth{}{October 2017\hb}

\author[]{\fnms{Kai} \snm{Chen}
	\thanks{Corresponding Author. E-mail: kai.chen@qub.ac.uk}},
\author[]{\fnms{Blesson} \snm{Varghese}},
\author[]{\fnms{Peter} \snm{Kilpatrick}},
and
\author[]{\fnms{Dimitrios S.} \snm{Nikolopoulos}}

\runningauthor{K. Chen et al.}
\address{School of Electronics, Electrical Engineering and Computer Science\\ Queen's University Belfast, UK}

\begin{abstract}
Existing power modelling research focuses not on the method used for developing models but rather on the model itself. This paper aims to develop a method for deploying power models on emerging processors that will be used, for example,  in cloud-edge data centers. Our research first develops a hardware counter selection method that appropriately selects counters most correlated to power on ARM and Intel processors. Then, we propose a two stage power model that works across multiple architectures. The key results are: (i) the automated hardware performance counter selection method achieves comparable selection to the manual selection methods reported in literature, and (ii) the two stage power model can predict dynamic power more accurately on both ARM and Intel processors when compared to classic power models. 

\end{abstract}

\end{frontmatter}

\section*{Introduction}
\label{sec:introduction}
Power monitoring has become a significant task for data-center management~\cite{ref1}. Direct power measurement obtained by physical meters or model-based interfaces has been widely supported in platforms~\cite{lrpm-3}. However, fine-grained power measurement of individual hardware/software components, which plays a significant role in runtime energy/performance management/optimisation, is not easy~\cite{lrpm-2}. For instance, both the model-based energy interface of the Intel Sandy Bridge server 
and the physical power meter of ARM Odroid-Xu3 board 
can measure the power of the entire processor rather than of individual computing cores in the processor or of the executing programs.

Developing accurate power models of computing cores or more fine-grained execution units, therefore, is an important avenue of research. A large proportion of existing models rely on multiple hardware activities of the processor represented by hardware performance counters, referred to as hardware counters, for estimating power~\cite{lrpm-3}.  

However, the hardware counters necessary to build an accurate power model may substantially differ across processors due to the differences in the instruction set, pipeline, cache architecture and on-chip interconnect. The hardware counters are usually selected on the basis of experimental knowledge of the processor~\cite{lrpm-2}. Such an approach used in traditional data centers cannot easily scale for heterogeneous processors (hosting multiple generations of server processors), or in emerging distributed computing environments like fog/edge computing and mobile cloud computing (in these environments, an application is distributed across data center processors, for example Intel Xeon processors~\cite{hcpm-xeon-1}, and low power processors, such as ARM~\cite{lrpm-arm-1} popularly employed in embedded systems for edge computing~\cite{offload-1}). In this paper, we make novel contributions by developing a hardware counter selection method that is employed across multiple processors without compromising the accuracy of the power model.

With hardware counters selected by our proposed method we evaluate three classic power models based on Linear Regression (LR), Support Vector Machines (SVM) and Neural Networks (NN) on Intel and ARM processors. 
This evaluation motivates the design of a two stage power model that takes advantage of LR to estimate a basic power value and then uses SVM to optimise this value for improved accuracy. It is observed that our model predicts dynamic power more accurately on both ARM and Intel processors when compared to classic power models.

The research contributions of this paper are: (i) the design and implementation of an automated hardware counter selection method to simplify the hardware counter selection process without sacrificing the accuracy of the power model, and (ii) the proposal of a Two Stage Power Model which takes advantage of both LR and SVM algorithms.

The remainder of this paper is organised as follows. 
Section~\ref{sec:definitions} presents the mathematical notation and the hardware platform employed. 
Section~\ref{sec:selectionmethod} proposes and evaluates a method for selecting hardware counters. 
Section~\ref{sec:twostagepowermodel} presents a two stage power model. 
Section~\ref{sec:conclusions} presents related work and concludes this paper.

\vspace{-0.25cm}
\section{Definitions}
\label{sec:definitions}
This section presents mathematical notation employed and the hardware platform used.

\textbf{\textit{Notations}}: Classic power models that are used for estimating dynamic power of processors and the concept of vectors are defined as used in this paper.  

\textit{Classic Power Models}: Consider a processor power model in which the estimated power, $P$, is the sum of the idle power consumed by the processor (static), and the power required for various activities of the processor (dynamic). Thus,
$P = P_{static} + P_{dynamic}$.
This paper explores dynamic power which is a function of the volume of hardware activities on the processor, obtained from a set of hardware counters.

Consider $n$ hardware counters whose values obtained from a processor during time interval $t_i$ are denoted as $e_{i_1}$, $e_{i_2}$, $\cdots$ , $e_{i_n}$ and let the measured dynamic power during interval $t_i$ be denoted by $P_{i_{dynamic}}$.  We consider the following three classic power models. 

\textit{a. Linear Regression Power Model (LRPM)}: In this model, dynamic power is defined as  
$P_{i_{dynamic}} = \sum\limits_{j=1}^n c_{j}e_{i_j}$, 
where $c_j$ is the coefficient of the $j^{th}$ hardware counter.

\textit{b. Neural Network Power Model (NNPM)}: Compared to LRPM which captures linear relationships, NNPM can also model non-linear relationships. Hardware counter values provided as input pass through layers of NNPM where a linear (i.e. the weighted sum) and non-linear function (i.e. activation function) are applied to map the input to the output.

\textit{c. Support Vector Machine Power Model (SVMPM)}:
This model captures both linear and non-linear relationships between dynamic power and the hardware counters. A set of hyperplanes are fitted using the training data to estimate the dynamic power. 

The ideal configuration of input parameters for both NNPM and SVMPM was manually chosen by extensively exploring the space. In this paper parameters that provide the most accurate estimation of dynamic power are chosen.  

\textit{Vectors}: We define vector 
$V_{i}~=~\{P_{i_{dynamic}}, e_{i_1}, e_{i_2}, \cdots , e_{i_n}\}$
where $P_{i_{dynamic}}$ and the $e_{i_j}$ are defined as above for the time interval $t_{i}$. 

Each vector is normalised to bring values of all variables into the same range between 0 and 1. The normalised vector of $V_{i}$ is represented as 
$\hat{V_{i}}~=~\{\hat{P}_{i_{dynamic}}, \hat{e}_{i_1}, \hat{e}_{i_2}, \cdots , \hat{e}_{i_n}\}$,
where 
$\hat{P}_{i_{dynamic}} =
P_{i_{dynamic}}$, $\hat{e}_{i_1} = \frac{e_{i_1}-min(e_1)}{max(e_1)-min(e_1)}$, $\cdots$ , $\hat{e}_{i_n} = \frac{e_{i_n}-min(e_n)}{max(e_n)-min(e_n)}$.

\textbf{\textit{Platform}}: Distributed computing environments such as those employed in fog/edge computing make use of both cloud data center and edge nodes. Typically, data center servers, for example Amazon cloud servers, make use of Intel Xeon processors\footnote{\url{https://aws.amazon.com/ec2/instance-types/}}, which are designed for high-performance computing. On the other hand edge nodes do not make use of large processors, instead employing low power processors, such as ARM\footnote{\url{http://www.arm.com/products/iot-solutions/mbed-iot-device-platform}}. Next generation power models will need to work for emerging distributed computing environments and therefore both an Intel Xeon processor representing servers used in data centers and an ARM processor representing edge nodes are used in our investigation. 

The first processor is the Intel Xeon Sandy Bridge server comprising two Intel Xeon E5-2650 processors with 8 cores on each processor.
The processor runs CentOS 6.5. We measure power consumption using the Running Average Power Limit (RAPL) interface~\cite{Intel-1}. 
The second processor is the ODROID-XU+E\footnote{\url{http://www.hardkernel.com}} board which has one ARM Big.LITTLE architecture Exynos 5 Octa processor. There are four Cortex-A15 cores and four Cortex-A7 cores and 2 GBytes of LPDDR3 DRAM. 
The system runs Ubuntu 14.04 LTS. 
We use the on-chip power meter to measure the power of the Cortex-A15 cores. 

The hardware counters are obtained by real-time profiling using Performance API (PAPI)
~\cite{PAPI-1}. Power is obtained from the on-chip power sensor on ARM and from the RAPL interface on Intel. This research employs 16 scientific benchmarks using MPI and OpenMP (such as the Buffon Laplace, Monte-Carlo and molecular dynamics simulations, solvers for Poisson and wave equations, and fast fourier transform) which captures a wide range of workloads\footnote{\url{http://people.sc.fsu.edu/~jburkardt/c_src/c_src.html}}. 
Vectors with hardware counters and measured power are continuously sampled during the execution of benchmarks approximately every 1 second. On ARM, we used the Cortex-A15 cores at their maximum frequency of 2.0GHz to execute the benchmarks. The Cortex-A7 cores at their maximum frequency of 1.4 GHz are used to obtain vectors. Similarly, on Intel, we used one processor at its maximum frequency to execute the benchmarks and the second processor at its maximum frequency to obtain vectors (both 2.0GHz).  

\vspace{-0.25cm}
\section{Hardware Counter Selection (HCS) Method}
\label{sec:selectionmethod}
\textbf{\textit{Design of the HCS method}}: 
To design a hardware counter based power model, the \textit{selection of hardware counters} must first be addressed. This requires addressing `how many' and `which' hardware counters should be selected.
 
If more hardware counters are employed, then a more accurate power model is built. Up to a maximum of 16 and 50 hardware counters are supported on ARM and Intel processors, respectively. However, using PAPI a maximum of only six hardware counters can be obtained simultaneously on both the ARM and Intel processors employed in this work. It should be noted that the number of hardware counters that are available and can be profiled simultaneously may vary between different processors even if they are produced by the same vendor. Multiplexing (profiling a different set of hardware counters sequentially to obtain a large number of hardware counters) techniques can surmount the limitation of the number of hardware counters that can be profiled, but introduces overheads that cannot be ignored. In this paper, we attempt to eliminate extra overheads by using only six hardware counters to build power models on both platforms, while retaining accuracy. Additional multiplexing techniques can easily be integrated into the modelling process proposed in this work.

In existing research, hardware counters that contribute to the power function are usually selected on the basis of experimental knowledge of the processor. Typically, all possible hardware counters that can be obtained are extensively explored using a cumbersome trial and error approach~\cite{lrpm-arm-1}. Then a combination of counters is chosen to develop a power model. However, this approach will not be practical for data centers in distributed cloud environments, such as in fog computing. It would be impossible to manually determine the suitable hardware counters of each processor for developing a power model. This motivates the need for the automated hardware selection method we propose.

We develop a generic Hardware Counter Selection (HCS) method that can be employed on multiple processors. The method selects a set of six hardware counters from all the available hardware counters that best correlates to power for a given processor. The method is based on a Random Forest (RF) algorithm that maps the hardware counters to power. We choose RF due to its accuracy in regression~\cite{RF-1}. We note that an RF based power model will not be feasible for on-line power monitoring due to its high computing complexity. However, the RF algorithm can determine and quantify the relative importance of each hardware counter to power during the model fitting process. Therefore, we leverage this characteristic of RF algorithms to build the HCS method that works off-line. 

The HCS method is designed to generate a list of hardware counters that are most relevant to power estimation. Algorithm~\ref{algo:selection} shows the proposed method. The key design principle is that the HCS method should be suitable for all applications and the hardware counters selected by the approach should not be dependent on a particular application. To obtain a general HCS method, we partition the dataset and obtain hardware counters for each subset to break any dependence on the dataset.
The inputs to the algorithm are:

1) $all\_vectors$, which is the set of all vectors (including all hardware counters from a processor) obtained from executing benchmarks using the multiplexing function of PAPI.

2) $n$, which is the number of hardware counters to be selected. In our case, we use six, which is the maximum number of counters obtained simultaneously using PAPI.

3) $ntree$, which is the number of trees that are used to build the random forest model for selection. This parameter is determined through an experimental exploration which estimates the effect of different values of $ntree$ on the hardware counter selection result; the experimental results are not within the scope of this paper. As a conclusion, the HCS method is not sensitive to $ntree$. In detail, employing values of $ntree$ which are not less than 2 on ARM and 16 on Intel, the HCS method selects the same hardware counters.

\begin{algorithm}
\caption{Hardware Counter Selection (HCS) Method}
\label{algo:selection}
\begin{algorithmic}[1]
\Procedure{select\_counters}{$all\_vectors$, $n$, $ntree$}
\State $counters\_selected \gets list\left(\right)$
\For{$i=1$ to $M$} \Comment{The entire dataset is partitioned into $M$ subsets }
   
   \State$part\_vectors \gets extract\left(all\_vectors, i, M\right)$ \Comment{Extract the $i^{th}$ subsets from}
   \Statex \hspace{22.5em} {overall M subsets}
   \State$rfes \gets $ randomForest$ \left(part\_vectors, ntree\right)$ 
   \State $counters\_importance[i] \gets rfes.importance$ 
   
\EndFor
\State Find n hardware counters with largest average value of importance. 

\State \textbf{Return} n events with largest average value of importance 
\EndProcedure
\end{algorithmic}
\end{algorithm}

The algorithm first partitions $all\_vectors$ into a set of subsets (i.e. $M$ subsets) (lines 3-4). During each run of the $for$ loop (line 3) for each subset $i$ which is extracted from the overall $M$ subsets (line 4), a Random Forest model is used to map hardware counters to power (line 5). The importance of each hardware counter for a given partition is obtained and stored in the $counters\_importance$ array (line 6). Finally, $n$ hardware counters with largest average importance values which are calculated by averaging the importance values of all subsets are found (line 7) and returned (line 8).

\textbf{\textit{Evaluation of the HCS method}}: 
To evaluate the Hardware Counter Selection (HCS) method, hardware counters selected by the HCS method and by the manual expert based method reported in the literature are compared. Then we compare the accuracy of classic power models presented in Section~\ref{sec:definitions}, when using hardware counters obtained from our selection method against a baseline using hardware counters reported in the literature. 

We use a rigorous training and testing strategy. All vectors obtained from profiling the execution of the benchmarks are partitioned equally into four parts. Then we use a combination of three parts to train the LRPM. The trained model is used to test: (i) vectors from the three parts used to train the model (75\% of the vectors), referred to as \textit{`Known'} vectors since they are known to the model through the training process; and (ii) vectors from the fourth part which were not used for training the model (25\% of the vectors), referred to as \textit{`Unknown'} vectors since they are not known to the model and were not used for training. 
  
Table~\ref{tab:hardwarecountersarm} and Table~\ref{tab:hardwarecountersintel} show the hardware counters reported in the literature (which we use as a baseline) and selected by the the HCS method for ARM and Intel, respectively. The baseline is determined by reviewing existing research~\cite{lrpm-arm-1, lrpm-arm-2, lrpm-4} and by considering the characteristics of our experimental platform and the profiling tool PAPI. 

\begin{table}[ht]
	\caption{Hardware counters from the baseline and selected by the HCS method on ARM}
	\label{tab:hardwarecountersarm}
\begin{center}   
	\begin{tabular}{c c l }
        \hline
        \multicolumn{2}{c}{\textbf{Hardware Counters}}  & \multirow{2}{*}{\textbf{Description}}  \\
        \cline{1-2}
        \textbf{Baseline} &  \textbf{HCS Method} \\
		\hline	
		\hline	
		\texttt{PAPI\_TOT\_CYC}	&  \texttt{PAPI\_TOT\_CYC}	&	Total cycles \\
		\texttt{PAPI\_TOT\_INS}	&  \texttt{PAPI\_TOT\_INS}	&	Instructions completed \\
        \texttt{PAPI\_TLB\_IM}	&  \texttt{PAPI\_TLB\_IM} & Instruction TLB misses\\
		\texttt{PAPI\_L1\_DCA}	&  \texttt{PAPI\_L1\_DCA}	&	L1 data cache accesses\\
		\texttt{PAPI\_L1\_ICA}	&  \texttt{PAPI\_L1\_ICA}	&	L1 instruction cache accesses\\
        \texttt{PAPI\_L2\_DCA}	&	- & Level 2 data cache accesses  \\
        - &  \texttt{PAPI\_L2\_TCM} & Level 2 cache misses \\
		\hline
	\end{tabular}
	\end{center}
\end{table}

\begin{table}[ht]
	\caption{Hardware counters from the baseline and selected by the HCS method on Intel}
	\label{tab:hardwarecountersintel}
\begin{center}   
	\begin{tabular}{c c l }
        \hline
        \multicolumn{2}{c}{\textbf{Hardware Counters}}  & \multirow{2}{*}{\textbf{Description}}  \\
        \cline{1-2}
        \textbf{Baseline} &  \textbf{HCS Method} \\
		\hline	
		\hline	
		\texttt{PAPI\_TOT\_CYC}	&	\texttt{PAPI\_TOT\_CYC}	&	Total cycles \\
		\texttt{PAPI\_TOT\_INS}	&	\texttt{PAPI\_TOT\_INS}	&	Instructions completed \\
		\texttt{PAPI\_LD\_INS}	&	\texttt{PAPI\_LD\_INS}	&	Load instructions\\
        \texttt{PAPI\_SR\_INS}	&	\texttt{PAPI\_SR\_INS}	&	Store instructions\\
		\texttt{PAPI\_FP\_OPS}	&	- & Floating point operations  \\
		\texttt{PAPI\_L3\_TCA}	&	- &	L3 total cache accesses \\
-	&	\texttt{PAPI\_REF\_CYC} & Reference clock cycles\\
        - & \texttt{PAPI\_L3\_TCM}	& L3 cache misses\\
		\hline
	\end{tabular}
	\end{center}
\end{table}

On the ARM and Intel processors we note that the hardware counters obtained from the HCS method are quite similar to those from the baseline (on ARM only one hardware counter is different and on Intel only two hardware counters differ). We infer from this that given different hardware processors our selection method can obtain appropriate hardware counters that capture dynamic power. It is also observed that the hardware counters for the ARM and Intel processors are different (4 out of the 6 hardware counters differ). The HCS method we propose can select processor dependent hardware counters that are suitable for developing power models. 

\begin{figure*}[ht]
\centering
	\subfloat[Known vectors]
	{	\label{knownvectorarm}
		\includegraphics[width=0.475\textwidth]
		{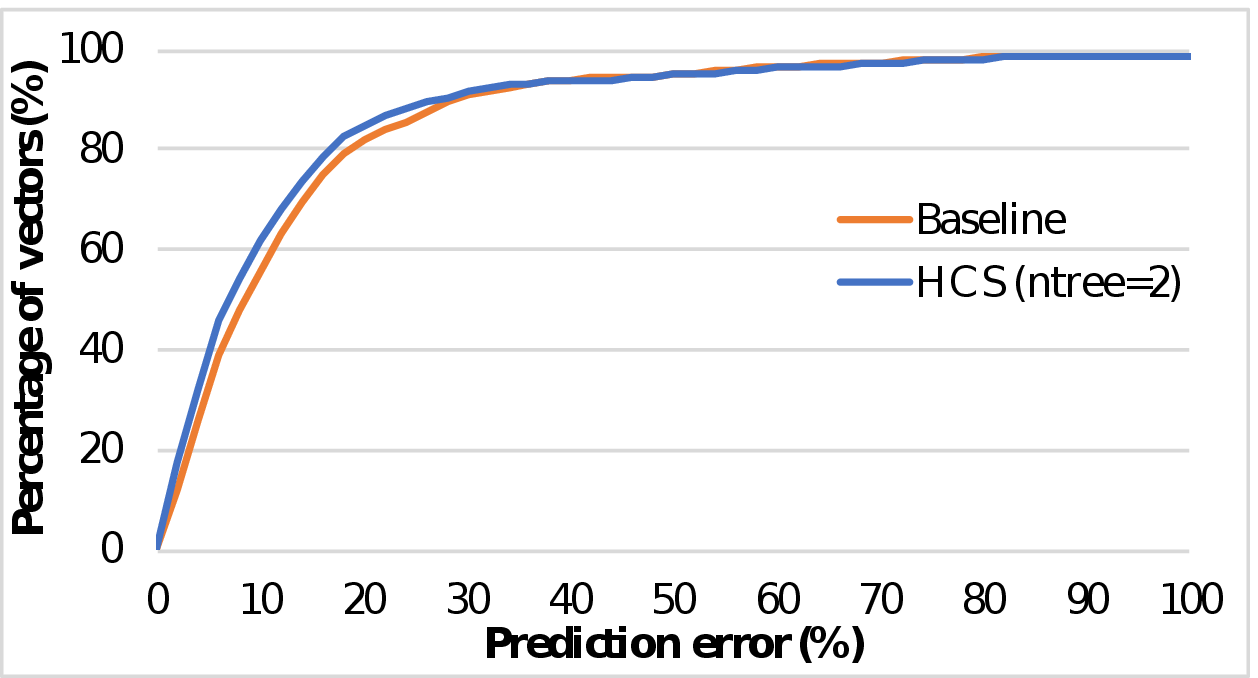}
	}
	\hfill
	\subfloat[Unknown vectors]
	{	\label{unknownvectorarm}
		\includegraphics[width=0.475\textwidth]
		{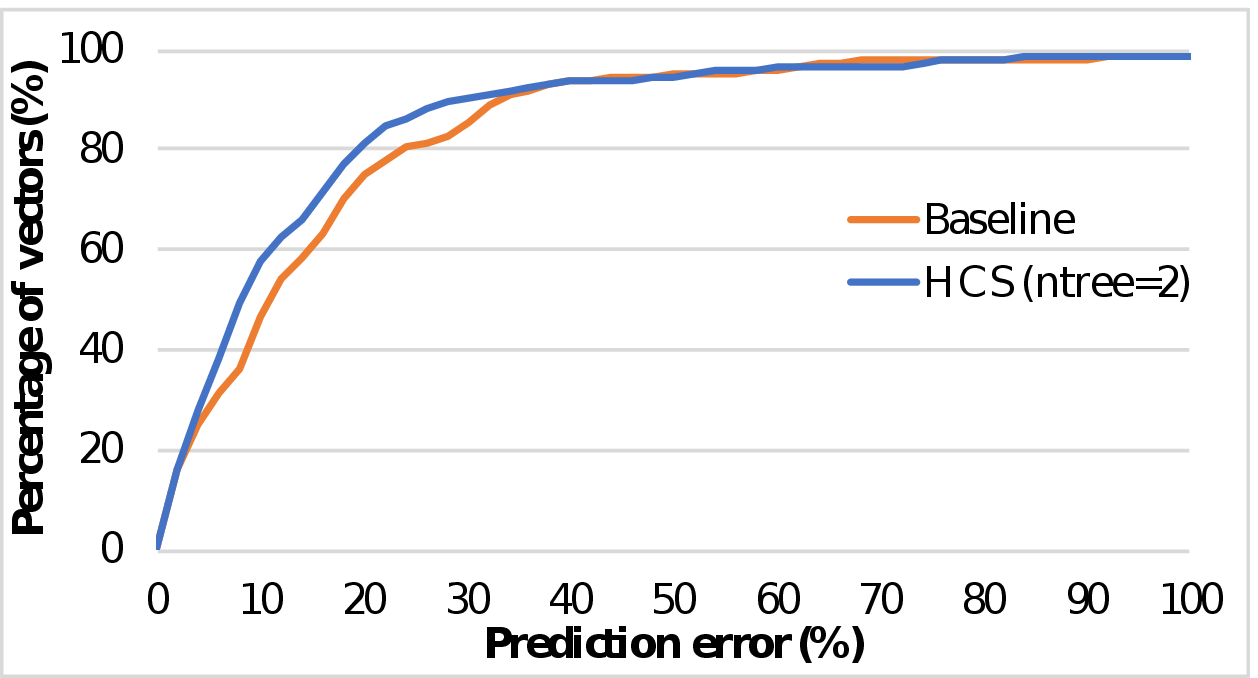}
	}\\
	\caption{Accuracy of the LRPM employing hardware counters reported in the literature referred to as baseline and selected by the HCS method ($ntree$ = 2) on the ARM processor}
	\label{fig:hardwarecounterselectorarm}
\end{figure*}

\begin{figure*}[ht]
\centering
	\subfloat[Known vectors]
	{	\label{knownvectorintel}
		\includegraphics[width=0.475\textwidth]
		{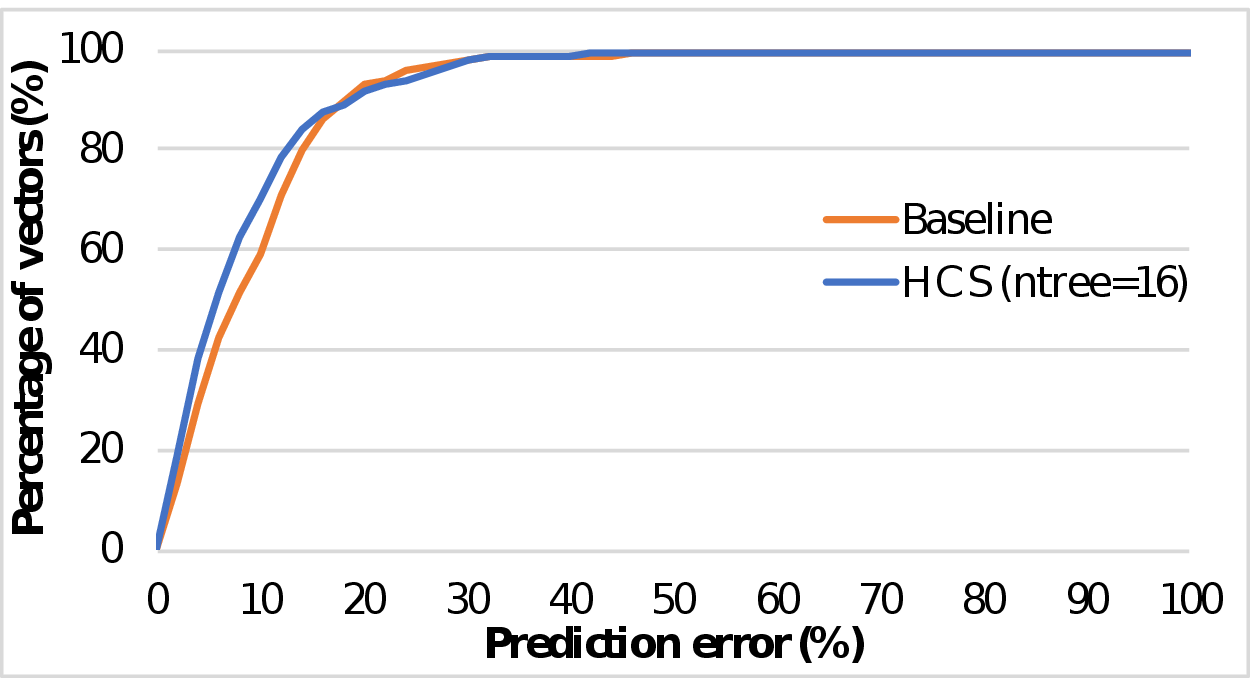}
	}
	\hfill
	\subfloat[Unknown vectors]
	{	\label{unknownvectorintel}
		\includegraphics[width=0.475\textwidth]
		{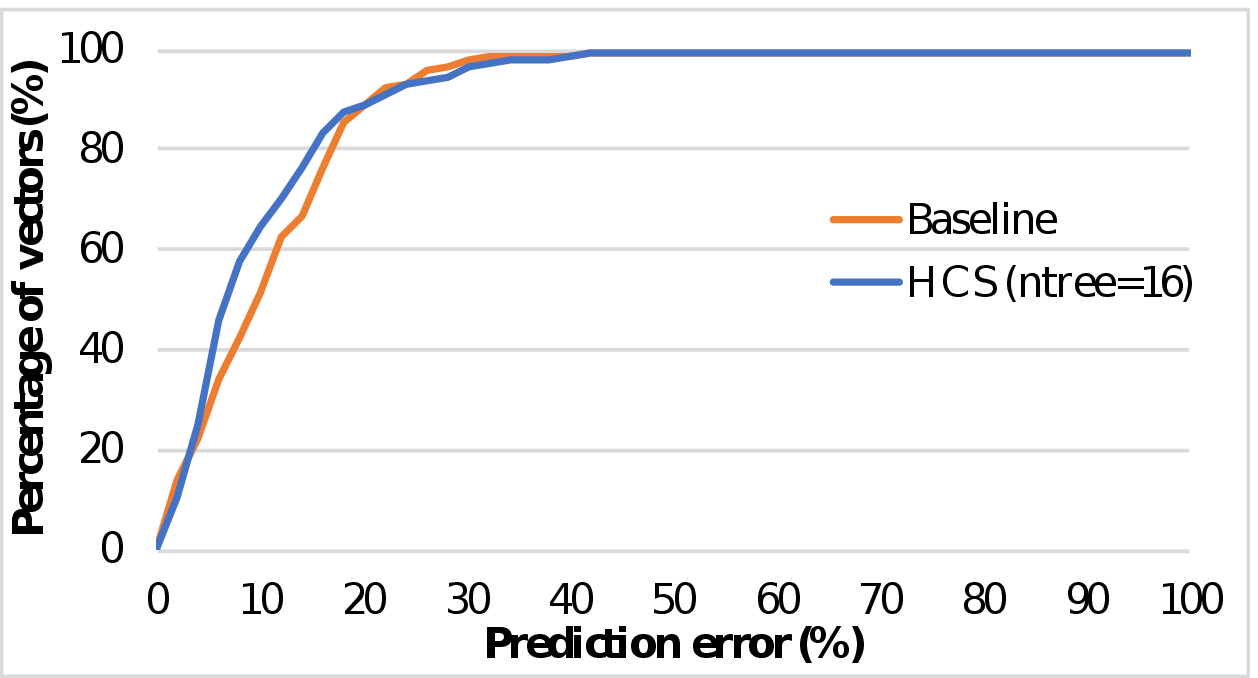}
	}\\
	\caption{Accuracy of the LRPM employing hardware counters reported in the literature referred to as baseline and selected by the HCS method ($ntree$ = 16) on the Intel processor}
	\label{fig:hardwarecounterselectorintel}
\end{figure*}

We evaluated the accuracy in predicting power based on estimation error $Error$, which is defined as 
$Error	= \frac{|P_{estimated} - P_{dynamic}|}{P_{dynamic}}$.
Figure~\ref{fig:hardwarecounterselectorarm} shows the percentage of Known and Unknown vectors that can be accurately predicted with different error percentages when using hardware counters of the baseline and the hardware counters from the HCS method on the ARM processor. Figure~\ref{fig:hardwarecounterselectorintel} corresponds to the Intel processor. In the best case, the HCS method on both processors performs better for Known and Unknown vectors than the baseline. The HCS method is automated in contrast to the baseline, but even in the worst case it provides near similar accuracy to the baseline. We evaluated the accuracy of power models based on Linear Regression (LRPM), Support Vector Machine (SVMPM) and Neural Network (NNPM). However, we present results based on LRPM, since the results and conclusions on all power models are similar.   

\vspace{-0.25cm}
\section{Design of a Two Stage Power Model}
\label{sec:twostagepowermodel}
In this section, we explore three classic power models to understand their accuracy. This exploration motivates the need for the Two Stage Power Model (TSPM). 

\textbf{\textit{Motivation}}:
We evaluate the three classic models by measuring the estimation error $Error$ in predicting power. The HCS method selects the hardware counters as shown in Table~\ref{tab:hardwarecountersarm} and Table~\ref{tab:hardwarecountersintel}. The training/testing strategy presented in Section~\ref{sec:selectionmethod} was used. 

\begin{figure*}
\centering
	\subfloat[On the ARM processor]
	{	\label{fig:comparisonclassicmodelsarm}
		\includegraphics[width=0.475\textwidth]
		{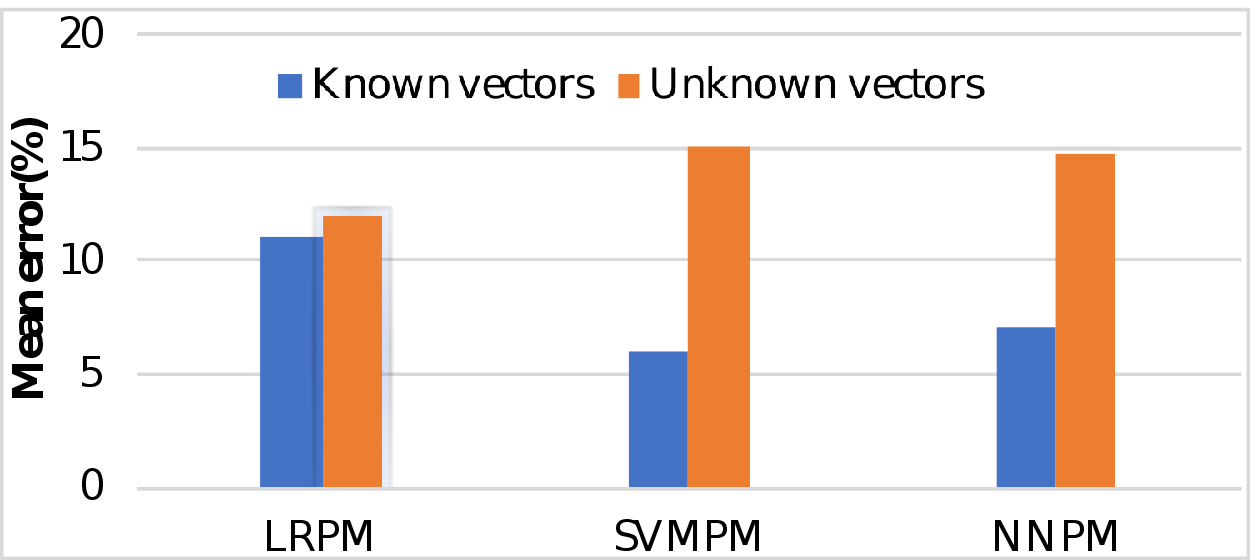}
	}
	\hfill
	\subfloat[On the Intel processor]
	{	\label{fig:comparisonclassicmodelsintel}
		\includegraphics[width=0.475\textwidth]
		{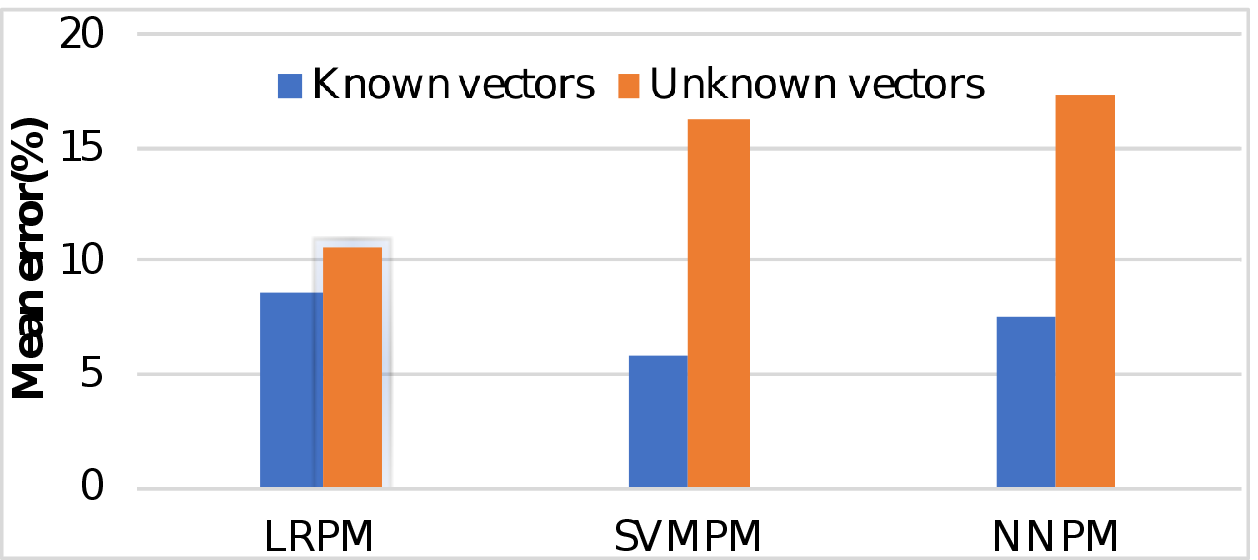}
	}\\
	\caption{The accuracy of classic power models for Known and Unknown vectors}
	\label{fig:comparison}
\end{figure*}

Figure~\ref{fig:comparison} shows the mean error of the classic power models for predicting dynamic power when testing Known and Unknown vectors on the ARM and Intel processors. On both processors, it is evident that the SVMPM is more accurate for predicting dynamic power of Known vectors. This indicates that SVMPM fits the training data well. Compared to SVMPM, LRPM relatively under fits the data resulting in lower accuracy. However, for Unknown vectors LRPM is most accurate. This is surprising, but is because more sophisticated models, such as SVMPM and NNPM over fit data~\cite{fitting-1} and lead to lower accuracy for Unknown vectors than a simpler model, such as LRPM. There is no off-the-shelf power model that achieves accuracy of the best performing power model for both Known and Unknown vectors. This motivates the need for a new power model that reduces the effect of over-fitted models in predicting Unknown vectors than classic power models, but at the same time achieves low error rates for known vectors. 

\textbf{\textit{Design}}:
In this section, we propose a Two Stage Power Model (TSPM) which takes advantage of the low variance of simple models, such as Linear Regression (LR) and of the low bias of sophisticated models, such as Support Vector Machine (SVM). We empirically identified that the combination of LR and SVM provided better accuracy than alternative classic models and was therefore chosen.

The TSPM operates in two stages. In the first stage, a LR based model is used to estimate a basic power value of an incoming vector. In the second stage, a SVM based model is employed to refine the basic power value to improve estimation accuracy. 

\begin{algorithm}
\caption{Training process of TSPM}
\label{trainfortspm}
\begin{algorithmic}[1]
\Procedure{Train\_Model}{$training\_vectors$}
\State $LRPM \gets build\_model\left(LR, training\_vectors\right)$ \Comment{LR is the abbreviation of Linear}  
\Statex \hspace{20.5em} {Regression}
\State $difference\_vectors \gets training\_vectors$ \Comment{Initialize the training set for the} \Statex \hspace{20.5em} {difference model (DM)}
\State $n \gets sizeof\left(training\_vectors\right)$ \Comment{n is the number of vectors in $training\_vectors$}
\For{$i=0$ to $n-1$}\Comment{Construct the training set for the difference model}
   \State $basic\_value \gets predict\left(LRPM, training\_vectors[i]\right)$
   \State $difference \gets training\_vectors[i,1] - basic\_value$
   \State $difference\_vectors[i,1] \gets difference$
\EndFor
\State $SVMDM \gets build\_model\left(SVM, difference\_vectors\right)$ 

\State \textbf{Return} $LRPM, SVMDM$
\EndProcedure

\end{algorithmic}
\end{algorithm}

\textit{Training of TSPM}: Algorithm \ref{trainfortspm} describes the training process of TSPM. First, an LRPM is developed using a training dataset consisting of profiled vectors as shown in Line 2. Then a difference based training dataset is constructed (Lines 3-8) by replacing the measured power of each vector in the original training set with the difference between the measured power and the value predicted by LRPM (Line 8). Finally, using the difference training set, a SVM based difference model is built (Line 9).

\textit{Prediction of TSPM}: Algorithm \ref{testfortspm} describes prediction of TSPM. For an incoming vector, both LRPM and SVMDM obtained from Algorithm~\ref{trainfortspm} are used. The LRPM is used to predict the basic power value (Line 2) and the SVMDM is used to estimate the difference between the measured power and the estimated power of LRPM (Line 3). We offset the basic power value with the difference, such that the final predicted power is obtained by summing the basic power and the difference (Line 4).

\begin{algorithm}
\caption{Prediction process of TSPM}
\label{testfortspm}
\begin{algorithmic}[1]
\Procedure{Predict\_power}{$test\_vector$, $LRPM$, $SVMDM$}
\State $basic\_value \gets predict\left(LRPM, test\_vector\right)$
\State $difference \gets predict\left(SVMPM, test\_vector\right)$
\State $power \gets basic\_value + difference$
\State \textbf{Return} $power$
\EndProcedure
\end{algorithmic}
\end{algorithm}

\begin{figure*}
\centering
	\subfloat[Prediction accuracy for Known vectors]
	{	\label{resknownarm}
		\includegraphics[width=0.475\textwidth]
		{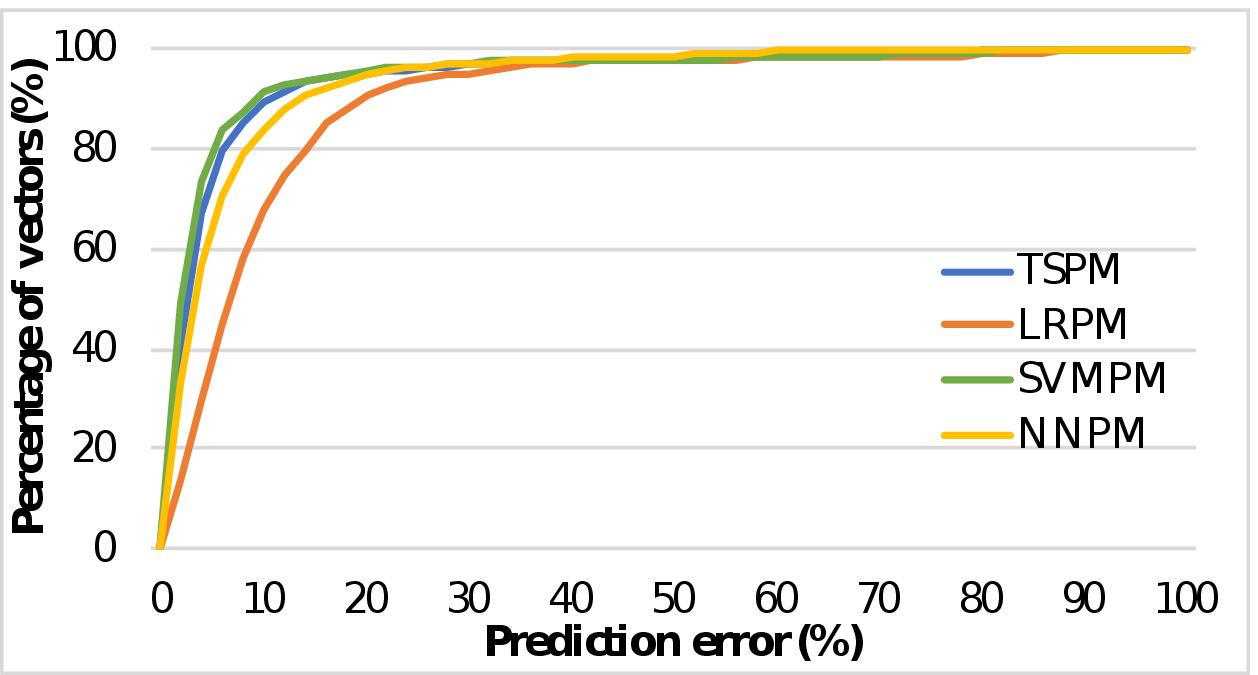}
	}
	\hfill
	\subfloat[Prediction accuracy for Unknown vectors]
	{	\label{resunknownarm}
		\includegraphics[width=0.475\textwidth]
		{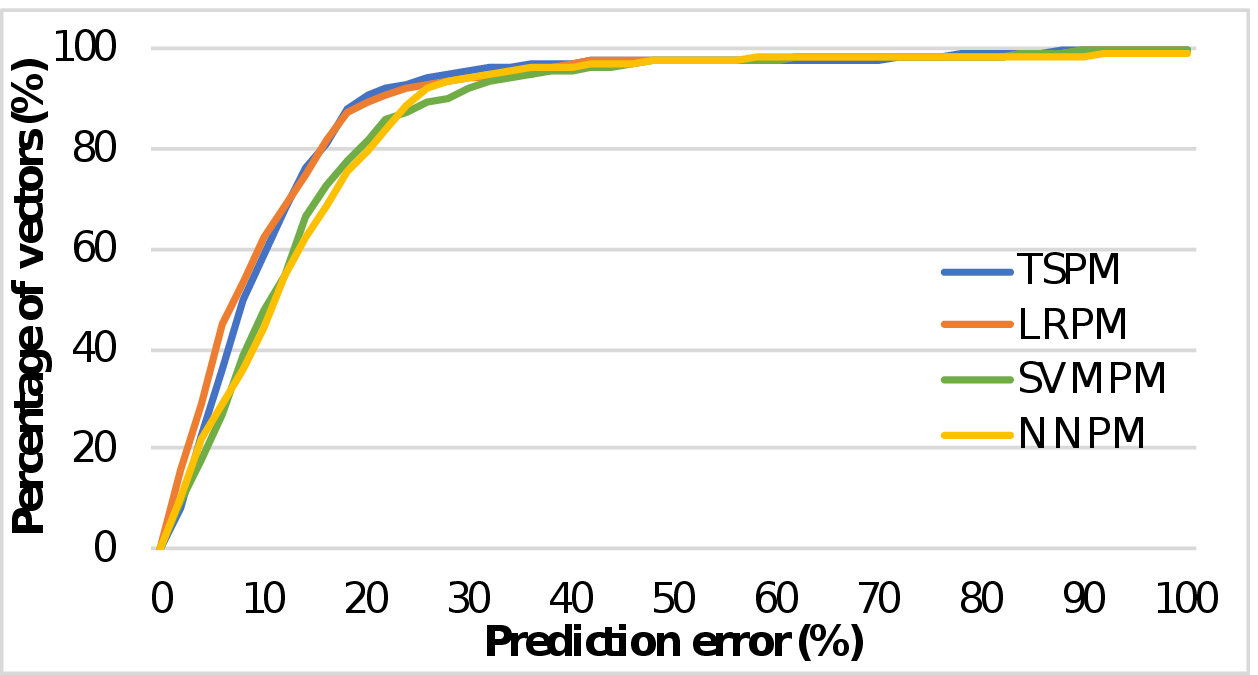}
	}\\
	\caption{Prediction accuracy of different models on the ARM processor}
	\label{resarm}
\end{figure*}

\textbf{\textit{Comparing TSPM and Classic Power Models}}:
In this section, the accuracy of the proposed TSPM against classic power models is evaluated. 
Figure~\ref{resarm} shows the prediction accuracy of TSPM in comparison to classic power models for both Known and Unknown vectors on ARM. For Known vectors, TSPM can achieve near similar accuracy to SVMPM (which has lowest prediction error for Known vectors). For Unknown vectors, TSPM obtains accuracy similar to the best classic model, which is LRPM when compared to SVMPM and NNPM. For example, nearly 60\% of Unknown vectors can be predicted with error less than 10\% using TSPM and LRPM.

\begin{figure*}
\centering
	\subfloat[Prediction accuracy for Known vectors]
	{	\label{resknownintel}
		\includegraphics[width=0.475\textwidth]
		{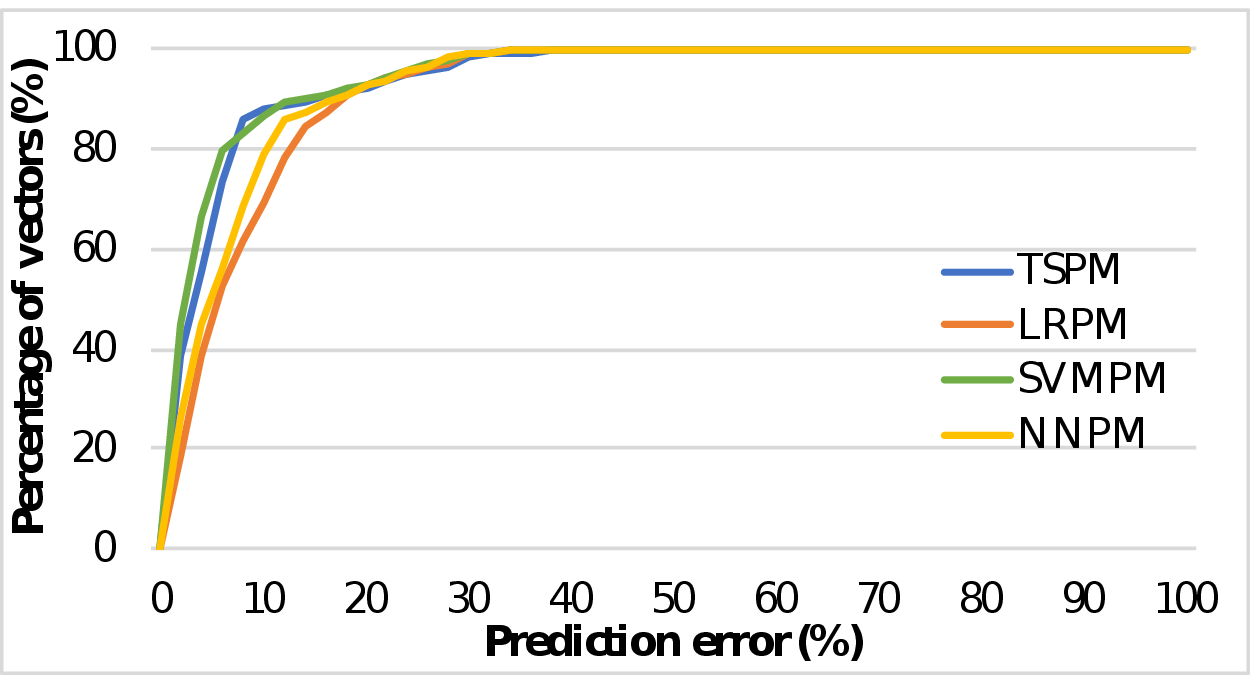}
	}
	\hfill
	\subfloat[Prediction accuracy for Unknown vectors]
	{	\label{resunknownintel}
		\includegraphics[width=0.475\textwidth]
		{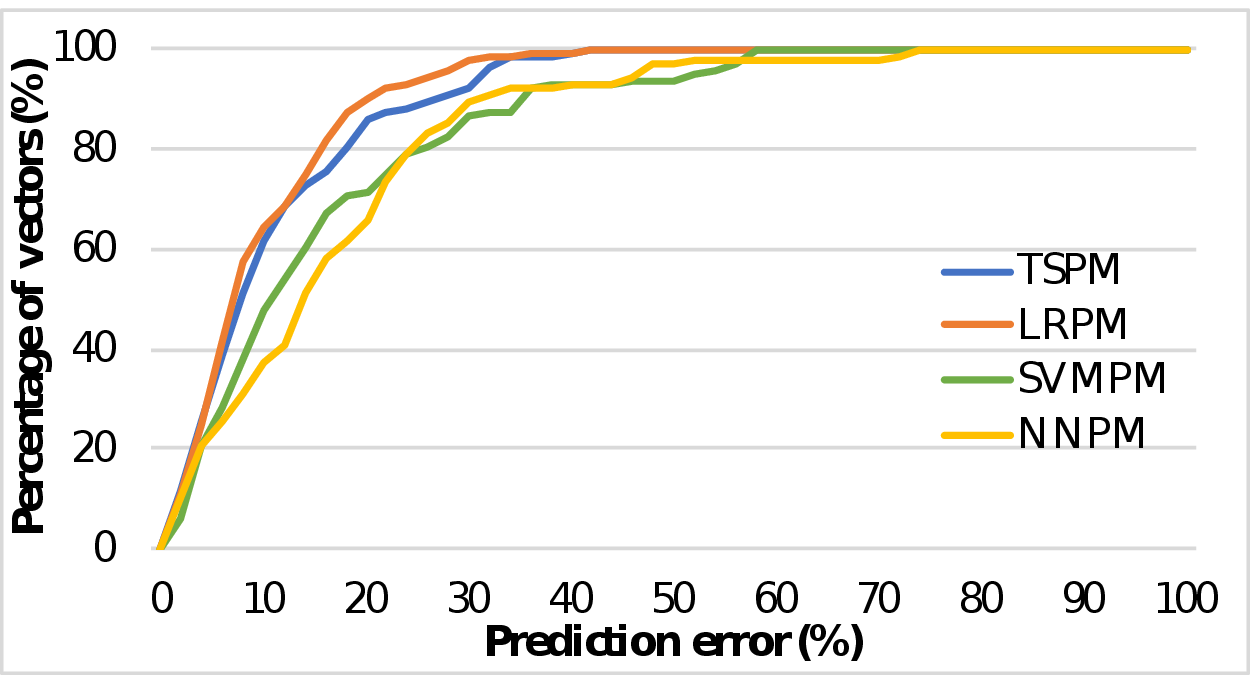}
	}\\
	\caption{Prediction accuracy of different models on the Intel processor}
	\label{resintel}
\end{figure*}

Figure \ref{resintel} shows prediction accuracy of TSPM when compared to classic power models for Known and Unknown vectors on Intel. On Intel for Known and Unknown vectors, TSPM performs similarly to the best classic power model. 

\vspace{-0.25cm}
\section{Discussion and Conclusions}
\label{sec:conclusions}
Research in power modelling has led to (i) instruction-level~\cite{instructionlevelmodel-2}, (ii) coarse-grain utilisation~\cite{utilisationlevelmodel-1}, and (iii) hardware counter-based~\cite{lrpm-3} models. 
Instruction-level models require extensive knowledge of the entire instruction set. It is cumbersome to obtain the power of each instruction and the overhead of all instruction pairs, thereby rendering these models impractical for real use. Although coarse-grained utilisation-based power models are easy to implement, they are not accurate since power depends not only on utilisation, but also on the type of operation. For example, floating point operations require more power than integer operations. Hardware counter-based power models are fine-grained utilisation models. These models are relatively simpler than instruction-level models, but at the same time are more accurate when compared to coarse-grained utilisation models. 

However, distributed computing models, such as fog/edge computing make use of processors in data centers and at the edge of the network. There are two challenges that will limit the use of existing power models in these settings. Firstly, hardware counter selection is usually dependent on human expertise. This becomes challenging when heterogeneous processors are used. Secondly, existing models focus on either large processors or edge-like processors, but do not work across both architectures. This limits the use of existing models for end-to-end power modelling since platform independent cross architectural models are required. Our research tackles both these challenges on multiple architectures by developing (i) an automated method for selecting hardware counters, and (ii) a two stage power model that performs better than existing models. 


The research in this paper firstly developed a hardware counter selection method that appropriately selects hardware counters that capture power for both ARM and Intel processors. This selection method simplifies the hardware counter selection process without compromising accuracy. Secondly, we developed a two stage power model that surmounts the challenges in using existing power models across multiple architectures. We demonstrated that our model predicts dynamic power more accurately on both ARM and Intel processors when compared to classic power models. 
\\
\\
\noindent \textbf{Acknowledgement} 
\noindent This research was funded by the SFI-DEL 14/IA/2474 grant.


\vspace{-0.5cm}
\bibliographystyle{IEEEtran}
\bibliography{references}

\end{document}